\documentclass[seceq]{ptptex}

\title{
Cosmological Constraints on a Massive Neutrino %
}

\author{
Masahiro \textsc{Kawasaki}$^{a,b}$ and Katsuhiko \textsc{Sato}$^{b,c}$%
}

\inst{
$^a$ Institute for Cosmic Ray Research,
     University of Tokyo, Chiba 277-8582, Japan\\
$^b$ Institute for the Physics and Mathematics of the Universe, 
     University of Tokyo, Chiba 277-8582, Japan\\
$^c$ Department of Physics, Meisei University, Hino, Tokyo 191-8506, Japan}

\abst{
The paper by Sato and Kobayashi in 1977 studied the cosmological 
effects of a massive neutrino and obtained  constraints on its 
properties. This paper initiated many studies to use cosmology 
as a laboratory of particle physics or to use particle physics to 
explore the very early universe. 
}

\begin{document}

\maketitle

\section{Introduction}
In 1977 Kobayashi and one of the present author (K.S.) studied 
cosmological effects of a massive neutrino and published 
the paper entitled {\it "Cosmological Constraints on the Mass
and the Number of Heavy Lepton Neutrinos"}~\cite{Sato:1977ye}.
Hereafter we call it Sato-Kobayashi 1977. 
This study was motivated by the discovery of anomalous lepton 
production in $e^{+}$-$e^{-}$ annihilation ($e^{+}+e^{-} \rightarrow 
e^{\pm}+\mu^{\mp}+{\rm missing~ energy}$) in 1975~\cite{Perl:1975bf},
which strongly suggested the existence of a heavy lepton 
having a mass in the range $1.6$ to $2.0$~GeV. [ Later, the existence of 
new lepton was confirmed and named tau ($\tau$). ]

Assuming the heavy lepton in the third family, the Sato-Kobayashi 1977
addressed  two issues; one was to constrain the mass of the associated 
neutrino and the other was
limiting the number of massive neutrinos. The latter issue had been 
partly investigated by Steigman et al.~\cite{Steigman:1977kc} 
for massless neutrinos and Kobayashi-Sato 1977 gave a more generic limit
which applies  to massive neutrinos. 
As for the first issue  the authors were first to study the cosmological 
effects of a massive neutrino and obtained  constraints on its mass and 
lifetime.  In deriving the cosmological constraints, the effects on
the cosmic matter density, cosmic background radiation and 
big bang nucleosynthesis were considered. Today this cosmological 
approach is the standard way to reveal properties of particles predicted 
in new models of particle physics or to explore the very early universe by
using new particle theories. The research field is now called particle 
cosmology.   
Therefore, Sato-Kobayashi 1977 is regarded as a historic paper that
took the initiative in development of particle cosmology. 

In this paper we review Sato-Kobayashi 1977 and the subsequent progress.
The paper is organized as follows. In section~\ref{sec:Sato-Kobayashi}
we   review Sato-Kobayashi 1977. In section~\ref{sec:progress}, the 
subsequent progress is shown, and we  conclude in section~\ref{sec:conclusion}.

\section{Review of Sato-Kobayashi 1977}
\label{sec:Sato-Kobayashi}

\subsection{Lifetime of the massive neutrino}

In the 1977 paper~\cite{Sato:1977ye}, a neutrino, which is associated with 
the heavy lepton ( now called `tau') suggested in the $e^{+}$-$e^{-}$ 
collider experiment, was 
considered. If it has a finite mass and qurk-like mixings with 
other neutrinos ($\nu_e, \nu_{\mu}$), the heavy neutrino is unstable 
and can decay into a lighter neutrino via
\begin{equation}
   \nu_h ~\rightarrow  \nu_{e,\mu} + \gamma .
   	\label{eq:decay_gamma}
\end{equation}
The neutrino with mass $m_{\nu_h} > 2m_e$ can also decay into 
an electron-positron pair as 
\begin{equation}
  \nu_h ~\rightarrow  \nu_e + e^{+} + e^{-}.
   	\label{eq:decay_ele}
\end{equation}
With use of the mass eigenstates $\nu_i~(i=1,2,3)$
the weak interaction eigenstate $\nu_h$ of the heavy neutrino is 
written as
\begin{equation}
   \nu_h = \sum_{i} U_{hi}\nu_i,
\end{equation}
where $U_{hi}$ is the neutrino mixing matrix. 
In Ref.\citen{Sato:1977ye},  the lifetimes for the
decay processes (\ref{eq:decay_gamma}) and (\ref{eq:decay_ele}) 
were estimated in the framework of Weinberg-Salam theory. 
For decay mode (\ref{eq:decay_gamma})
\begin{eqnarray}
   \tau(\nu_h ~\rightarrow  \nu_{e(\mu)} + \gamma) 
   & = & \frac{128\pi^3}{G_F^2\alpha m_{\nu_h}^5}\frac{16}{9}
   \left(\sum_i U_{ih}U^*_{i e(\mu)}\frac{m_i^2}{m_W^2}\right)^{-2}
   \nonumber  \\
   & \simeq & 2\times 10^{6} {\rm yr} \left(\frac{m_{\nu_h}}{\rm MeV}
   \right)^{-5}
   | U_{he(\mu)}|^{-2},
   \label{eq:lifetime_gamma}
\end{eqnarray}
where $m_{\nu_h}$ is the mass of the heavy neutrino, $m_i$ is the charged 
lepton mass, $\alpha$ is the fine structure constant, $G_F$ is the Fermi 
constant and $M_W$ is the $W$ boson mass.
In the second line of Eq.(\ref{eq:lifetime_gamma}) 
we have used the measured $\tau$ lepton mass, but in 1977, the heavy
lepton ($= \tau $) was not confirmed, so the upper bound on the lifetime
was obtained from unitarity of the mixing matrix, which leads to
the following inequality:
\begin{eqnarray}
     \sum_{i' < i}(\sum_{j}U_{ji}U^*_{ji'}m_j^2)^2
     & \le & \sum_{j}U_{ij}U^*_{ij}m_j^4 
     -(\sum_{j}U_{ji}U^*_{ji}m_j^2)^2  \nonumber \\
     & \le & \frac{1}{4} \max_{j,j'}[(m_j^2- m_{j'}^2)^2].
\end{eqnarray}
Using this relation and $m_{j}/M_W < 1$, the lower bound on the 
lifetime was obtained as
\begin{equation}
   \label{eq:lifetime_gamma_SK}
   \tau(\nu_h ~\rightarrow  \nu_{e(\mu)} + \gamma) 
    >  \frac{128\pi^3}{G_F^2\alpha m_{\nu_h}^5}\frac{64}{9}
   \left(\frac{\max_j[m_j^4]}{m_W^4}\right)^{-2}
   > 2~{\rm yr} \left(\frac{m_{\nu_h}}{\rm MeV}\right)^{-5}.
\end{equation}

On the other hand, the lifetime for decay into $e^{+}~e^{-}$ is given by
\begin{eqnarray}
   \tau(\nu_h ~\rightarrow  \nu_{e} + e^{+} + e^{-}) 
    & = & \frac{192\pi^3}{G_F^2 m_{\nu_h}^5} |U_{1h}U^*_{1e}|^{-2}
   \nonumber  \\
    & \simeq & 3\times 10^{4}~ {\rm sec} \left(\frac{m_{\nu_h}}{\rm MeV}\right)^{-5}
   | U_{h1}|^{-2},
   \label{eq:lifetime_ele}
\end{eqnarray}
for $m_{\nu_h} > 2m_e$. The mixing  $U_{1h}$ is restricted with use of 
$\mu$-$e$ universality. The pion decay process $\pi \rightarrow e+\nu_h$,
if exists, is enhanced because the amplitude is proportional to $m_{\nu_h}$. 
So in Ref.~\citen{Sato:1977ye} requiring that the contribution of
$\pi \rightarrow e+\nu_h$  to  the total decay width of
$\pi \rightarrow e+\nu$ should be less than 5\%, the
limit on the $U_{1h}$ was estimated as  
\begin{equation}
   |U_{1h}|^2(m_{\nu_h}/m_e)^2 < 0.05,
\end{equation}
which led to the lower limit to the lifetime,
\begin{equation}
   \label{eq:lifetime_ele_SK}
   \tau(\nu_h ~\rightarrow  \nu_{e} + e^{+} + e^{-}) 
    \gtrsim  2\times 10^{8}~ {\rm sec} 
    \left(\frac{m_{\nu_h}}{\rm MeV}\right)^{-3}.
\end{equation}

Eqs.(\ref{eq:lifetime_gamma_SK}) and
(\ref{eq:lifetime_ele_SK}) show that the massive 
neutrino decays with cosmological time scale and the 
decay into an electron-positron pair, if kinematically possible, 
may be faster than that into a photon.  

\subsection{Cosmological constraints}

Since the massive neutrino is long-lived ($\tau > 1$~sec), it 
can affect the thermal history of the universe. 
In Sato-Kobayashi~1977~\cite{Sato:1977ye} 
the  constraints on mass and lifetime 
were obtained by considering the following cosmological effects: 

\vspace{0.5cm}
\begin{enumerate}
\item When the massive neutrino is stable or has a longer lifetime 
than the present age of the universe, its present mass density 
may exceed the observed upper limit.
\item Photons and electrons produced in the decay may cause 
a spectral distortion of the cosmic microwave background (CMB) 
radiation, or 
if the decay takes place after recombination, the emitted
photons may be directly observe as the background radiation. 
\item The massive neutrino can affect the big bang nucleosynthesis 
(BBN) by speed-up of the cosmic expansion and/or entropy production. 
\end{enumerate}
\vspace{0.5cm}

First let us consider the cosmic density of the heavy neutrino. 
The neutrinos are in thermal equilibrium via weak interaction
($\nu_h + \bar{\nu}_h \leftrightarrow e^{-} + e^{-}$)
in the very early universe. The annihilation cross section is
\begin{equation}
    \langle \sigma v \rangle \sim \left\{ 
      \begin{array}{ll}
         G_F^2 T^2 &  m_{\nu_h} \lesssim T \\[0.5em]
         G_F^2 m_{\nu_h}^2 &  m_{\nu_h} \gtrsim  T
      \end{array}\right. , 
\end{equation}
where $v$ is the relative velocity and $\langle\cdots\rangle$ means 
thermal average.  
When the interaction rate $\Gamma = \langle \sigma v \rangle n_\nu$   
becomes less than the cosmic expansion rate $H \sim T^2/M_p$  
($M_p$: Planck mass $\simeq 2.4\times 10^{18}$~GeV), the neutrino
decouples from the thermal bath.  After decoupling, the neutrino 
does not interact with other particles and its comoving density  
freezes  out. The freezing temperature is determined from 
$\Gamma \simeq H$, leading to 
$T_f \simeq \max[1, (m_{\nu_h}/10{\rm MeV})]$~MeV. 
Then the number ratio of the massive 
neutrino to the photon is estimated as 
\begin{equation}
      \left(\frac{n_{\nu_h}}{n_{\gamma}}\right)
      \simeq \frac{3}{11} 
      \min \left\{1, \left(\frac{10~{\rm MeV}}{m_{\nu_h}}\right)^3
      \right\}.         
\end{equation}
Notice that the neutrino number density decreases for 
$m_{\nu_h} \gtrsim 10$~MeV
because such heavy neutrino becomes non-relativistic at decoupling 
and its number density suffers from the Boltzmann suppression
($\sim \exp(-m/T)$). 

\subsubsection{Stable Neurino}

If the massive neutrino is stable or has a longer lifetime than 
the age of the universe, the neutrino density should be less than 
the upper limit of the present density of the universe. 
The present mass density of the heavy 
neutrino is given by 
\begin{equation}
      \rho_{\nu_h} =
      \frac{3}{11}g_s m_{\nu_h}n_{\gamma,0}
      \min \left\{1, \left(\frac{10~{\rm MeV}}{m_{\nu_h}}\right)^3
      \right\},        
\end{equation}
where $g_s$ is the statistical weight of the massive 
neutrino.\footnote{%
The statistical weight $g_s$ is always $1$ for a Majorana neutrino. 
As for a Dirac neutrino $g_s$ is $1-2$ depending on its mass.
Because of chirality of the weak interaction, the rate of 
the interaction of a right-handed neutrino is suppressed by
a factor $(m_{\nu}/E)^2$ where $E$ is the relevant energy. 
Thus, neutrinos with mass less than $2$~keV are never in thermal 
equilibrium in the early universe and $g_s=1$ for that 
case~\cite{Dolgov:2002wy}. 
However, in Ref.\citen{Sato:1977ye}
$g_s=2$ was assumed for any massive neutrinos. }
In term of the density parameter $\Omega \equiv \rho/\rho_c$,
where $\rho_c$ is the critical density 
($= 1.054\times 10^4 h^2$~eV/cm$^3$) and $h$ is the Hubble constant
in units of $100$km/s/Mpc, the above equation is rewritten as
\begin{equation}
      \Omega_{\nu_h} = 1.0~h^{-2}~g_s 
      \left(\frac{m_{\nu_h}}{100{\rm eV}}\right)
      \min \left\{1, \left(\frac{10~{\rm MeV}}{m_{\nu_h}}\right)^3
      \right\}.
\end{equation}
In Ref.\citen{Sato:1977ye}, from observations of the deceleration 
parameter [$q_0 = (\ddot{a}a/\dot{a}^2)$] and the age of the 
universe at that time, $\Omega h^2 < 1.45$ was adopted as 
the upper limit of the density parameter, 
which resulted in 
\begin{equation}
    m_{\nu_h} < 70~{\rm eV} ~~~~{\rm or}~~~~
    m_{\nu_h} > 3.7~{\rm GeV} .
\end{equation}
As mentioned before, since the Boltzmann suppression reduces the 
number density of the heavy neutrino, the neutrino with  
large mass $m_{\nu_h} \gtrsim 3$~GeV is allowed as well as the light 
neutrino with mass less than $O(10)$~eV. 
Unfortunately, the lower bound on the heavy neutrino mass is 
often called ``Lee-Weinberg limit'' because Lee and 
Weinberg~\cite{Lee:1977ua} obtained the limit at the same 
time.\footnote{%
The both papers~\cite{Sato:1977ye,Lee:1977ua} were submitted in May 1977. 
We should also note that the lower bound on the heavy neutrino mass 
was also obtained  by P.Hut~\cite{Hut:1977zn},
Dicus, Kolb and Teplitz~\cite{Dicus:1977nn} and Vysotskii, 
Zel'dovich and Dolgov~\cite{Vysotsky:1977pe}. 
}

\subsubsection{Unstable Neutrino}
 
Next, we consider the constraint on the unstable massive neutrino 
which decays into a photon or an electron-positron pair. 
If the neutrino decays sufficiently early $ t \lesssim 10^{10}$~sec, 
the emitted photons are quickly thermalized through
Compton scattering and bremsstrahlung. However, when decay takes 
place at later epochs, the emitted photons may partially thermalized 
via Compton scattering but cannot form the Planck distribution 
because photon number changing processes such as bremsstrahlung 
are inefficient. Thus, the CMB spectral shape is distorted, which 
leads to a constraint on the lifetime $\tau_{\nu_h}$
of the massive neutrino. 

When photons are emitted in the neutrino decay, 
the ratio of the extra photon density $\Delta \rho_\gamma$ 
to the CMB density $\rho_\gamma$   at the decay time
is estimated as
\begin{equation}
	 \label{eq:decay_photon_density}
     \frac{\Delta\rho_\gamma(\tau_{\nu_h})}{\rho_{\gamma}(\tau_{\nu_h})}
      = \frac{(3/11)g_s (m_{\nu_h}/2)n_{\gamma}F}
      {2.7T_d n_{\gamma}} \simeq 0.05g_s \frac{m_{\nu_h}}{T_d}F,
\end{equation}
where $T_d$ is the cosmic temperature at $t=\tau_{\nu_h}$ and 
$F = \min[1, (10~{\rm MeV}/m_{\nu_h})^3]$. With relation between the 
cosmic temperature and time, $t = 10^{12}~{\rm sec} (T/{\rm eV})^{-3/2}$,
and $\Delta \rho_{\gamma}/\rho_\gamma \lesssim 0.1$, the constraint was 
obtained~\cite{Sato:1977ye} as
\begin{equation}
       \tau_{\nu_h} \lesssim \left\{
       \begin{array}{ll}
           10^{12}~{\rm sec}  
           \left(\frac{m_{\nu_h}}{\rm eV}\right)^{-3/2} 
           &  m_{\nu_h} \lesssim 20~{\rm eV}  
           \\[0.5em]
           10^{10}~{\rm sec}
           &  20~{\rm eV} \lesssim m_{\nu_h} \lesssim 2~{\rm GeV}
           \\[0.5em]
           10^{9}~{\rm sec}\left(\frac{m_{\nu_h}}{\rm GeV}\right)^{3}
           & m_{\nu_h} \gtrsim 2~{\rm GeV}
       \end{array}\right.  .
\end{equation}
The above constraint obtained for decay into photons also applies
to the case where electron-positron pairs are produced in the decay,
i.e, for $m_{\nu_h} > 2m_e$ because the energetic electrons
(positrons) scatter off the CMB photons to produce energetic photons 
or the positrons annihilate
into  photons. 

When the decay takes place after recombination the emitted photons 
freely stream and may be observed as the background radiation 
after suffering redshift. This 
extra background radiation should not exceed the observed one, from which 
Sato-Kobayashi~1977 obtained the constraint on the lifetime of the 
massive neutrino as 
\begin{equation}
  \tau_{\nu_h} \gtrsim 10^{18}~{\rm sec}
\end{equation}

Finally, let us show how Sato-Kobayashi~1977 derived the 
constraint on the lifetime and mass of the massive neutrino 
from consideration of nucleosynthesis in the early universe. 
Two effects both of which increase the 
$^4$He abundance were studied in Ref.\citen{Sato:1977ye}. 

The first effect is entropy production by the neutrino decay. 
It is well known that the abundances of light elements synthesized
in the standard BBN only depend on the baryon density of the 
universe which is parametrized by ``photon-baryon ratio'',
\begin{equation}
    \eta_B \equiv \frac{n_{B}}{n_{\gamma}}
    = 2.7\times 10^{-8}\Omega_B h^2.
\end{equation}
In Ref.\citen{Sato:1977ye} the different parameter $h_N= \rho_B/T_9^3$
($T_9=T/10^9$K) was used but here we have adopted the standard 
parameter used in recent literature. $h_N$ is related to $\eta_B$ as
\begin{equation}
    h_N = 3.4\times 10^{-6} (\eta_B/10^{-10}) {\rm g/cm}^{3}.
\end{equation}
When the neutrino decay produces many photons, the baryon-photon 
ratio at BBN epoch ($\eta_{\rm BBN}$) is  larger than 
the present one and is roughly given by
\begin{equation}
     \eta_{\rm bbn} \simeq 
     \left(1+ \frac{\Delta\rho_\gamma(\tau_{\nu_h})}
     {\rho_{\gamma}(\tau_{\nu_h})}\right)
     \eta_{B 0},
\end{equation}
where $\Delta\rho_\gamma/\rho_{\gamma}$ is given by 
(\ref{eq:decay_photon_density}).
Using the lower bound on the present baryon density ($\Omega_B > 0.05$), 
which predicts the lowest $^4$He abundance 
because the theoretical prediction of $^4$He abundance is an
increasing function of $\eta_{\rm bbn}$, $\eta_{\rm bbn}$ is estimated
as
\begin{equation}
	\label{eq:eta_bbn}
    \eta_{\rm bbn} \lesssim 3.5\times 10^{-10}
         \left(1+\frac{\Delta\rho_\gamma(\tau_{\nu_h})}
     {\rho_{\gamma}(\tau_{\nu_h})}\right).
\end{equation}

The other effect is speed-up of the universe due to the massive
neutrino. As compared with massless case, the massive neutrino 
can have a larger energy density, which increases the expansion rate
of the universe and leads to earlier freeze-out of neutron-proton
exchange interactions (e.g. $n + \nu_e \leftrightarrow p +e^{-}$). 
Since the neutron-proton ratio at the  freeze-out temperature $T_f$
is given by $n/p \simeq \exp(-1.29{\rm MeV}/T_f)$, the earlier 
freeze-out (higher $T_f$) results in more abundance of $n$ and
hence more production of $^4$He. The change of the cosmic expansion
rate is represented by $\xi$ defined as
\begin{equation}
    \xi \equiv \left(\frac{\rho(T)}{\rho_{\rm std}(T)}\right)^{1/2},
\end{equation}
where $\rho_{\rm std}$ is the cosmic density at temperature $T$
for the standard case without massive neutrinos. 
The densities $\rho_{\rm std}$ and $\rho$ are given by
\begin{eqnarray}
    \rho_{\rm std}  =  \frac{\pi^2}{30} 
    \left(\frac{11}{2} T^4
    + \frac{7}{4}N_{\rm std} T_{\nu}^4
    \right), ~~~~~~~
    \rho   \simeq   \rho_{\rm std} 
    + (m_{\nu_h} + 3.15T_{\nu})n_{\nu_h},
\end{eqnarray}
where $T_{\nu}$ and $N_{\rm std}$ are the neutrino temperature 
and the number of massless neutrino species. 
As mentioned before, in 1977  the existence of the tau lepton was
not confirmed, thus $N_{\rm std} = 2$ was taken as the standard 
value. Since the speed-up effect is important when the neutron-proton
ratio freezes out ($T_f \simeq 1$~MeV.), we can assume $T >m_e$ and
hence $T_{\nu} = T$. Thus $\xi$ is written as
\begin{equation}
   \label{eq:xi_bbn}
   \xi^2 = 1 + 0.062 g_s\left(\frac{m_{\nu_h}}{T} + 3.15\right).
\end{equation}
On the other hand the theoretical prediction for the $^4$He abundance
was calculated numerically by Wagoner~\cite{Wagoner:1972jh} 
and he gave the empirical formula, 
\begin{equation}
   \label{eq:He4_th}
   Y = 0.264 + 0.0195\log (\eta_{\rm bbn}/10^{-10}) + 0.380\log \xi,   
\end{equation}
where $Y$ is the mass fraction of $^4$He, $Y=\rho_{^4{\rm He}}/\rho_B$.
Using Eqs.(\ref{eq:eta_bbn}), (\ref{eq:xi_bbn}) and (\ref{eq:He4_th})
together with observational upperbound $Y < 0.29$, the following
constraint on the mass and lifetime was obtained~\cite{Sato:1977ye}:
\begin{equation}
	\label{eq:bbn_cons_SK}
    \tau_{\nu_h} \lesssim \left\{
    \begin{array}{ll}
       1~\sec \left(\frac{m_{\nu_h}}{\rm MeV} \right)^{-2}
       &  (m_{\nu_h} \lesssim 1~{\rm MeV})\\[0.5em]
       1~\sec \left(\frac{m_{\nu_h}}{\rm MeV} \right)^{4}
       &  (m_{\nu_h} \gtrsim 1~{\rm MeV})
    \end{array}\right.         
\end{equation}


The cosmological constraints obtained in Ref.\citen{Sato:1977ye} were
so stringent that a large region in the parameter space 
$(\tau_{nu_h}, m_{\nu_h})$ was excluded as shown in Fig.4 
of Ref.\citen{Sato:1977ye}. Furthermore, together with 
the lower limits on the lifetime of the massive neutrino 
estimated in the Weinberg-Salam theory (\ref{eq:lifetime_gamma_SK}) and
(\ref{eq:lifetime_ele_SK}), the neutrino in the mass range,
\begin{equation}
   70~{\rm eV} ~< ~ m_{\nu_h}~ < ~ 23~{\rm MeV}
\end{equation}
was found to be excluded. 

\subsection{Number of Massive Neutrinos}

Furthermore, Sato-Kobayashi~1977 discussed the limit on 
the number of massive neutrinos. As seen in the previous subsection,
the masses of neutrinos should be less than $70$~eV 
or more than $23$~MeV. So we can assume that $N_1$ types of neutrinos
have masses less than $70$~eV and $N_2$ types of neutrinos
have masses larger than $23$~MeV. 

The constraint on $N_1$ is obtained by considering the speed-up 
effect in BBN.  In this case, assuming massless $\nu_e$ and $\nu_\mu$,
Eq.(\ref{eq:xi_bbn}) is rewritten as
\begin{equation}
   \xi^2 = 1+ 0.196~g_s N_1.
\end{equation}
Using Eq.(\ref{eq:He4_th}) and $Y < 0.29$ the constraint is obtained as
\begin{equation}
   N_1 \le 3~ g_{s}^{-1}.
\end{equation}

As for $N_2$, suppose that $\nu_m$ has the
longest lifetime among neutrinos with mass larger than $23$~MeV.
Then the lifetime is bounded as
\begin{equation}
  \tau_{\nu_m} \gtrsim     
  \frac{192\pi^3}{G_F^2}\left(\frac{1}{N_2}
  \sum_i m_{\nu_i}^5 |U_{1i}|^2\right)^{-1}
  \gtrsim \frac{192\pi^3}{G_F^2 m_{\nu_{N_2}}^3 m_e^2}
  \left(\frac{1}{N_2}
  \sum_i  |U_{1i}|^2(m_{\nu_{i}}/m_e)^2\right)^{-1},
\end{equation}
where $\nu_{N_2}$ is the neutrino with the largest mass. 
Using $\mu$-$e$ universality Sato-Kobayashi~1977 obtained
\begin{equation}
    \tau_{\nu_m} \gtrsim  2.2\times 10^6 N_2 
    (m_{\nu_{N_2}}/{\rm MeV})^{-3}~{\rm sec}.
\end{equation}
In Ref.\citen{Sato:1977ye}, applying the BBN constraint, $N_2$ was
limited as
\begin{equation}
   N_2 ~\lesssim ~  
   \left(\frac{m_{\nu_{N_2}}}{23~{\rm MeV}}\right)^7.
\end{equation}

\section{Progress after 1977}
\label{sec:progress}

Since Sato and Kobayashi first obtained the constraint on the massive 
neutrino in 1977, the constraints have been improved according 
to development of cosmology, laboratory experiments and observations. 

\subsection{Stable Neutrino}

The neutrino, if stable, cannot 
have a mass larger than about $70$~eV from the cosmic density constraint.
In other words, the neutrino can be dark matter of the universe 
if its mass is $O(10)$~eV. 
In fact, the neutrino with $O(10)$~eV attracted much interest in 
cosmology in 1980's.
In addition, at that time it was reported that a Russian group~\cite{Lubimov:1980}
discovered the electric neutrino mass in the 
tritium beta decay ($^3{\rm H} \rightarrow ^3{\rm He} + e^{-}
+\bar{\nu}_e$) experiment [ which was later denied by other experiments].  
However, it was soon found that a large velocity dispersion of the 
neutrino erases the small scale density fluctuations by
free-streaming. So if the neutrino is dark matter,
very large structure such as clusters of galaxies are formed first 
and smaller structure like galaxies are formed later. 
This ``top-down'' scenario was studied extensively by $N$-body 
simulations and it was concluded that the neutrino cannot be a 
main component of the dark matter~\cite{Blumenthal:1987cr}. 

Here, before discussing the recent cosmological bound on the neutrino 
mass, we briefly describe the experimental progress towards 
measurement of neutrino masses.
The experimental evidence for massive neutrinos was first
obtained by the Super-Kamiokande(SK) experiment~\cite{Fukuda:1998mi} 
which measured atmospheric neutrinos and discovered  
the neutrino oscillation between $\mu$- and $\tau$-neutrinos.
The experiment suggests 
\begin{equation}
    | m_{\nu_\tau}^2 -  m_{\nu_\mu}^2 | 
    \simeq 3\times 10^{-3}~{\rm eV}^2.
\end{equation}
Moreover, the solar neutrino 
experiments~\cite{Fukuda:2001nj,Ahmad:2002jz} and the reactor 
experiment~\cite{Eguchi:2002dm} 
also see the oscillation between $e$- and $\mu$- neutrinos and obtained 
\begin{equation}
    | m_{\nu_\mu}^2 -  m_{\nu_e}^2 | 
    \simeq 7\times 10^{-5}~{\rm eV}^2.
\end{equation}
Notice that the oscillation experiments can measure only the
differences of neutrino mass squares and do not give the 
absolute values. The most severe mass limit is still obtained
by the tritium beta decay experiment. The present best limit 
is~\cite{Amsler:2008zzb} 
\begin{equation}
    m_{\nu_e} ~< 2~{\rm eV}.
\end{equation}
As for the other neutrinos, the present mass bounds are~\cite{Amsler:2008zzb}
\begin{eqnarray}
   \label{eq:nu_mu_mass}	
   m_{\nu_\mu} & < & 0.19~{\rm MeV} \\
   \label{eq:nu_tau_mass}	
   m_{\nu_\tau} & < & 18.2~{\rm MeV} 
\end{eqnarray}

Let us return to cosmological consideration. 
Even if the massive neutrino is a minor component of the dark
matter, its existence can give a significant effect on the 
density fluctuations on small scales. Fukugita, Liu and 
Sugiyama~\cite{Fukugita:1999as},
comparing the predicted density fluctuations with that inferred from 
the cluster abundance, derived the neutrino 
mass limit as
\begin{equation}
     \sum m_{\nu_i} \lesssim 3~{\rm eV},
\end{equation}
where the sum is taken over the three species of neutrinos. 
The result of the oscillation experiments strongly suggest that
the three neutrino masses are degenerate if their masses are larger than
$0.1$~eV, i.e. 
$m_{\nu_e}\simeq m_{\nu_\mu}\simeq m_{\nu_\tau}$.

The neutrino mass limit was significantly improved by Wilkinson 
Microwave Anisotropy Probe (WMAP) observation which  obtained 
the  full sky map of the CMB temperature and thereby determined the cosmological 
parameters precisely~\cite{Spergel:2003cb}. The neutrinos with mass 
$\gtrsim O(0.1)$~eV become non-relativistic before or during 
recombination and they change the epoch of matter-radiation equality, 
which alters the amplitude of the temperature fluctuations through 
change of the gravitational potential from radiation dominated era 
to matter dominated one.  The  recent WMAP
five year data alone gives the upper bound on the neutrino mass 
as~\cite{Komatsu:2008hk}
\begin{equation}
       \sum m_{\nu_i}  < 1.7~{\rm eV} ~~~~(95\% {\rm CL}),
\end{equation}
and a more stringent upper bound is obtained if 
the other observational data (baryon acoustic oscillation 
and supernovae ) are added~~\cite{Komatsu:2008hk}, 
\begin{equation}
       \sum m_{\nu_i}  < 0.61~{\rm eV} ~~~~(95\% {\rm CL}).
\end{equation}

\subsection{Unstable Neutrino}

In the 1977 paper, the cosmological constraints on the radiatively 
decaying  neutrino were obtained by considering effects of the 
emitted photons on the background radiation and big bang 
nucleosynthesis. Even today these effects lead to the most stringent 
constraint on long-lived radiatively decaying particles 
including massive neutrino. 

\subsubsection{CMB spectral distortion}

As for the effect on the background radiation, 
the emitted photons (or electron-positron pairs) cause the spectral
distortion of CMB as pointed out in Sato-Kobayashi 1977\cite{Sato:1977ye}. 
If the neutrino decays early enough the photons injected in the decay 
are fully thermalized by photon number violating processes.  
In Ref.\citen{Sato:1977ye}, the bremsstrahlung 
($e + p \rightarrow e + p + \gamma$) was considered 
as the main photon number violating process. Afterwards, 
double Compton scattering ($e + \gamma \rightarrow e + 2\gamma$) 
was recognized to thermalize the background
photons more efficiently by Lightman~\cite{Lightman::1981}.
If we include the double Compton scattering, it is found that 
the injected photons are  thermalized and the Planck 
distribution is recovered quickly for $t \lesssim 10^6$~sec. 

However, photons produced  between 
decoupling of double Compton scattering ($10^{6}$~sec) 
and recombination ($10^{13}$~sec) may cause the spectral 
distortion in CMB. When the energy exchange between 
electrons and photons occurs sufficiently
frequently by Compton scatterings ($t < 10^8$~sec), 
the spectrum of CMB photons becomes Bose-Einstein distribution 
with chemical potential $\mu$.  
On the other hand, if the Compton scattering is less frequent
($t > 10^9$~sec) the spectrum distortion is described 
by $y$ parameter.@
The COBE satellite launched in 1989 showed that the CMB spectrum
is that of a nearly perfect blackbody, from which the distortion
parameters $\mu$ and $y$ are stringently 
constrained as~\cite{Fixsen:1996nj}
\begin{equation}
   |\mu | < 9\times 10^{-5}~~~~y < 1.2\times 10^{-6}.
\end{equation}
These distortion parameters are related to the energy injection 
$\Delta \rho_{\gamma}$ as 
\begin{equation}
  \frac{\Delta\rho_{\gamma}}{\rho_{\gamma}} \simeq 
  \left\{\begin{array}{ll}
   0.714\mu   &
   (10^6~{\rm sec}\lesssim \tau_{\nu_h} \lesssim 10^{9}~{\sec})\\[0.5em]
    4y        &
   (10^9~{\rm sec}\lesssim \tau_{\nu_h} \lesssim 10^{13}~{\sec})
   \end{array}\right.  ,
\end{equation}
where $\Delta\rho_{\gamma}/\rho_{\gamma}$ is given by 
Eq.(\ref{eq:decay_photon_density}) ( a factor $(4/3)$ should be 
multiplied for decay into an electron-positron pair).
Thus we can obtain the following constraint: 
\begin{equation}
   \label{eq:cmb_dist_const}
   \tau_{\nu_h} \lesssim \left\{
   \begin{array}{ll} 
      10^3~{\rm sec} \left(\frac{m_{\nu_h}}{\rm eV}\right)
      & (m_{\nu_h} \lesssim 1~{\rm keV}) \\[0.6em]
      10^6~{\rm sec}
      & ( 1~{\rm keV}\lesssim  m_{\nu_h}\lesssim 30~{\rm MeV})\\[0.6em]
      10^{12}~{\rm sec} \left(\frac{m_{\nu_h}}{\rm GeV}\right)^4
      & (m_{\nu_h} \gtrsim 30~{\rm MeV})
   \end{array}
   \right.  ,      
\end{equation}
where we have taken account of ``time dilation'', i.e. 
$\tau_{\rm eff}$(effective lifetime)$ \simeq (T_d/m_{\nu_h})\tau_{\nu_h}$ in the small 
neutrino mass range~\cite{Miyama:1978mn,Kawasaki:1985ff}. 

\subsubsection{Big Bang Nucleosynthesis}

The BBN constraint discussed in Sato-Kobayashi~1977 was rather 
rough one and the more precise calculation was done by Miyama and 
one of the present author (KS)~\cite{Miyama:1978mn}, which was later
updated in Ref.\citen{Terasawa:1988my}.  In those studies, 
the abundances of the light elements (D, $^3$He, $^4$He, $^7$Li ) were
estimated by calculation of the nuclear reaction network, and 
the result was compared with those inferred  from observations.    
The resultant constraint is approximately given by 
\begin{equation}
    \label{eq:bbn_const}
    \tau_{\nu_h} \lesssim \left\{
    \begin{array}{ll}
       3\times 10^2~\sec \left(\frac{m_{\nu_h}}{\rm MeV} \right)^{-2}
       &  ( m_{\nu_h} \lesssim 2~{\rm MeV})\\[0.5em]
       10^2~\sec 
       &  (2~{\rm MeV} \lesssim m_{\nu_h} \lesssim 20~{\rm MeV})
    \end{array}\right.         
\end{equation}

The above constraint is imposed mainly from entropy
production due to the radiative decay of the massive neutrino.
However, Lindley~\cite{Lindley:1984bg} pointed out that 
the radiative decay of the neutrino with mass larger 
than $10$~MeV has another serious effect on BBN.
High energy photons with $O(10)$~MeV produced in the decay 
can destroy the light elements
which are synthesized in BBN through the following reactions:
\begin{eqnarray}
   \label{eq:He4_a}
   ^4{\rm He} + \gamma  & ~\rightarrow ~&  {\rm T} + p  
   ~~~~~~~~(19.8~{\rm MeV}),  \\
   \label{eq:He4_b}
   ^4{\rm He} + \gamma  & ~\rightarrow ~ &  ^3{\rm He} + n  
   ~~~~~(20.6~{\rm MeV}) , \\
   \label{eq:He3}
   ^3{\rm He} + \gamma  & ~\rightarrow ~ &  {\rm D} + p  
   ~~~~~~~~(5.5~{\rm MeV}) , \\
   \label{eq:D}
   {\rm D} + \gamma  & ~\rightarrow ~ &  n + p  
   ~~~~~~~~~(2.2~{\rm MeV}) , \\
   & \ldots &    \nonumber
\end{eqnarray}
where the numbers in parentheses denote the threshold energies. 
High energy photons and/or electrons injected into the cosmic plasma
induce electro-magnetic 
showers and quickly thermalized, and how fast  photons 
lose their energy crucially depends on whether 
the photon energy $E_\gamma$ is high enough to scatter off
the background photon to produce electron-positron pairs
($\gamma + \gamma_{\rm BG} \rightarrow e^{-}+e^{+}$). 
For pair creation to occur both energies of the background photon 
and the photon in the shower should be large, which leads to 
the condition,
\begin{equation}
   \label{eq:phptpn-photon}
   E_{\gamma} \ge E_{\rm th} \equiv \frac{m_e^2}{22T}.
\end{equation}
Thus, high energy photons with $E_\gamma > E_{\rm th}$ can be thermalized
by the photon-photon process so quickly that they have little chance
to destroy the light elements. On the other hand, photons with lower 
energy cannot partake in the photon-photon process and are only 
thermalized by much slower Compton scatterings. Such photons can 
destroy the light element if their energy is above the thresholds 
of the photo-dissociation reactions.  
Using Eqs.(\ref{eq:He4_a})--(\ref{eq:phptpn-photon})
it is found that when the heavy neutrino
decays at $t \gtrsim 10^4 (10^6)$~sec, the photons in the induced 
electromagnetic showers can destroy D ($^4$He).
This photo-dissociation effect was studied in details for radiative decay of 
massive neutrinos in Refs.\citen{Kawasaki:1986sn,Terasawa:1988my},
which leads to the following constraint:
\begin{equation}
     \label{eq:photo_dis_const}
     \tau_{\nu_h} \lesssim 10^{4}~{\rm sec} ~~~~~~~
     (10~{\rm MeV} \lesssim m_{\nu_h} \lesssim 1~{\rm GeV}).
\end{equation}

\subsubsection{Constraint from Supernova 1987A}

So far we have reviewed the progress of the cosmological
constraints on the radiatively decaying neutrino from the
CMB distortion and BBN. Other astrophysical and experimental 
constraints are also discussed in the literature. Among these 
constraints, here, we present the one from the supernova 1987A. 
SN1987A is a historic supernova since the neutrinos emitted its core
were observed first in the human history. 

As well known, the neutrinos emitted from the supernova carry 
almost all explosion energy ( $\sim 10^{53}$~erg) which is much 
larger than is visible ($\sim 19^{47}$~erg). Thus, if 
even a small fraction of the neutrinos
decay into photons (or electrons) inside the envelop of the progenitor 
( $\lesssim 3\times 10^{12}$~cm ), the visible luminosity of the 
supernova is significantly changed, from which the stringent 
constraint on the lifetime is obtained as~\cite{Kolb:1990vq}
\begin{equation}
   \label{eq:SN_const}
   \tau_{\nu_h}  \gtrsim 
   \left\{
   \begin{array}{ll}
       10^7~{\rm sec} \left(\frac{m_{\nu_h}}{\rm MeV} \right) 
       &   (m_{\nu_h} \lesssim 10~{\rm MeV})\\[0.6em]
       4\times 10^7~{\rm sec}
       \left(\frac{m_{\nu_h}}{\rm MeV} \right)^{3/2}
       \exp\left(-\frac{m_{\nu_h}}{\rm MeV}\right)
       &   (m_{\nu_h} \gtrsim  10~{\rm MeV})
   \end{array}\right.  .
\end{equation}

One can see from Eqs.~(\ref{eq:nu_mu_mass}),
(\ref{eq:nu_tau_mass}	), (\ref{eq:cmb_dist_const}),
(\ref{eq:bbn_const}), (\ref{eq:photo_dis_const})
and (\ref{eq:SN_const}) that the radiative
decay of the massive neutrino is almost forbidden.

\subsection{Non-radiative Decay}  

In some models a massive neutrino can decay non-radiatively.
For example, in the majoron (familon) model where a global lepton 
number (family) symmetry 
is spontaneously broken and associated Nambu-Goldstone boson called
a majoron (familon) appears, a neutrino can decay into a lighter 
neutrino and a  majoron (familon). Non-radiative decay was studied 
in Ref.\citen{Terasawa:1987pg}, and more recently 
in Ref.\citen{Kawasaki:1993gz}. From these studies, 
it is known  that 
non-radiative decay is also stringently constrained by cosmology. 

\subsection{Number of Neutrinos}

The number of neutrino species $N_{\nu}$ was determined by measuring 
the decay width of the $Z$ boson, which leads to~\cite{Amsler:2008zzb} 
\begin{equation}
    N_{\nu} = 2.9840 \pm 0.0082.
\end{equation}
Therefore we now know that the only three types of neutrinos exist
in nature.

\section{Conclusion}
\label{sec:conclusion}

We have seen that the 1977 paper by Sato and Kobayashi~\cite{Sato:1977ye}
is considered as a historic paper which first showed that valuable information
on neutrino properties can be obtained from cosmological consideration. 
The obtained constraints on radiative decay of a massive neutrino were 
updated by a number of authors and led to the important conclusion that
the neutrino radiative decay is almost forbidden. 
The value of the paper is not limited in neutrino physics. The arguments
also apply to more generic particles that decay on cosmological time scales
as the subsequent studies showed.  

The paper initiated many studies to use cosmology as a laboratory of 
particle physics or to use particle physics to explore the very 
early universe. 
Today this research field is called particle cosmology and has achieved 
brilliant success. 
For example, the idea of inflation~\cite{Sato:1980yn,Guth:1980zm} 
was born during effort 
to make monopole in the grand unified theories compatible with cosmology. 
Another example is interplay between supersymmetry and cosmology. 
Theories based on supersymmetry (SUSY) are considered as 
the most promising candidate for models beyond the standard particle 
physics,  and predict many new particle some of which have long lifetimes.
One of such long-lived particles is stable and account for the dark matter 
of the universe~\cite{Jungman:1995df}, 
while others often cause serious cosmological problems like
gravitino problem~\cite{Kawasaki:2004qu}. 
In any way,  it is now  the standard procedure to test 
particle physics model by cosmology. 

In concluding this article, we would 
like to quote the final sentence of Ref.\citen{Sato:1977ye},
\vspace{0.5cm}
\begin{quote}
   {\it It is very interesting that valuable information on particle 
    physics can be derived from cosmological arguments in spite of 
    large uncertainties inherited there.\\
    \hfill K. Sato and M. Kobayashi (1977)}
\end{quote}
\vspace{0.5cm}

\section*{Acknowledgements}
The work is supported in part by World Premier International Research
Center Initiative, MEXT, Japan,@and@by Grants-in-Aid for Scientific
Research provided by the Ministry of Education, Science and Culture of
Japan through Research Grants S19104006 (K.S.) and No.14102004 (M.K.).

%

\end{document}